\documentstyle[prl,aps,epsf,twocolumn]{revtex}
\begin{document}
\draft
\preprint{}

\title{ Large anisotropy in the optical conductivity  of YNi$_2$B$_2$C
}
\author{S. J. Youn$^1$, B. I. Min$^2$, and A. J. Freeman$^3$ }
\address{
       $^1$National Creative Research Initiative Center for Superfunctional
       Materials and Department of Chemistry,
       Pohang University of Science and Technology, Pohang 790-784, Korea.
}
\address{
$^2$ Department of Physics,
       Pohang University of Science and Technology, Pohang 790-784, Korea.
}
\address{
$^3$
Department of Physics and Astronomy
Northwestern University, Evanston, IL 60208-3112, USA
}


\maketitle
\begin{abstract}
The optical properties of YNi$_2$B$_2$C are studied by using
the first-principles
full-potential linearized augmented plane wave (FLAPW) method within the
local density approximation.  Anisotropic behavior is obtained in
the optical conductivity, even though the electronic
structure shows 3D character.
A large peak in $\sigma_z$ is obtained at 2.4 eV.
The anisotropic optical properties are analyzed
in terms of interband transitions between energy levels and found
that the Ni site plays an important role.
The electronic energy loss spectroscopy (EELS) spectra are also
calculated to help elucidate the anisotropic properties in this system.
\end{abstract}

Quaternary transition metal borocarbide superconductors
with formula RM$_2$B$_2$C (R= Y, rare-earths; M=transition metal)
have attracted much attention due to their relatively high
superconducting transition temperatures in the intermetallic compounds
including exchange enhanced transition metal elements such as
Ni, Pd, and Ru \cite{nagarajan,cava}.
Of special interest is the coexistence and interplay between superconductivity
and magnetism in the systems containing rare-earth elements\cite{canfield}.
YNi$_2$B$_2$C with $T_c$=15.6 K serves as a reference system
for understanding the superconducting mechanism
in this family, because there are no complications introduced by the presence
of magnetic rare-earth elements.

YNi$_2$B$_2$C crystallizes in a body-centered tetragonal structure with
a ThCr$_2$Si$_2$-type (space group $I4/mmm$, $D^{17}_{4h}$).
The crystal structure is layer-like in the $c$ axis,
reminiscent of the high-$T_c$ cuprate superconductors,
consisting of two alternating layers of Ni$_2$B$_2$ and YC.
Electronic structure studies show three dimensional (3D)
character and so it is thought to be a conventional BCS type
superconductor with a relatively high density of states (DOS) at the
Fermi level \cite{mattheiss,pickett,jilee,singh}.
Despite the layered anisotropic crystal structure, some physical
properties of YNi$_2$B$_2$C are often considered to be isotropic.
Note that the 3D character of the electronic structure
does not necessarily mean the isotropy of physical properties.
In fact, it has not yet been settled whether the physical properties
of YNi$_2$B$_2$C  are isotropic or not.
Civale {\it et al.} \cite{civale} observed an anisotropic effective mass ratio
$\gamma=(m_{max}/m_{min})^{1/2}\sim 1.1$, while
Johnston-Halperin {\it et al.} \cite{johnston} observed an isotropic
result,
$\gamma\sim 1.005$.
Fisher {\it et al.} \cite{fisher} reported isotropic resistivity for
non-magnetic YNi$_2$B$_2$C and anisotropic resistivity for magnetic
borocarbide superconductors like ErNi$_2$B$_2$C and HoNi$_2$B$_2$C.
Rathnayaka {\it et al.} \cite{rathnayaka} observed an isotropic upper
critical field, $H_{c2}$, for YNi$_2$B$_2$C, but a
small anisotropic $H_{c2}$ for LaNi$_2$B$_2$C.
In contrast, anisotropic properties were reported
for YNi$_2$B$_2$C thin-films in the upper critical field\cite{vaglio},
thermal conductivity\cite{sera},
critical field for vortex lattice transition \cite{sakata},
and paramagnetic susceptibility from nuclear magnetic resonance (NMR)
experiments\cite{kumagai}.
von Lips {\it et al.} \cite{vonlips} used X-ray absorption
spectroscopy (XAS) and the linear combination of atomic-like orbitals
(LCAO) band method for YNi$_2$B$_2$C to determine the anisotropic nature
of the unoccupied DOS.

\begin{figure}[b]
     \epsfysize=7cm
     \epsfbox{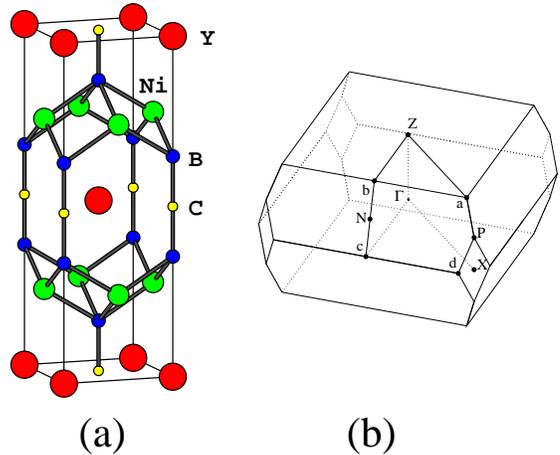}
     \caption{
     (a) Crystal structure and
     (b) the first Brillouin zone of YNi$_2$B$_2$C.
       \label{fig.bz}
     }
\end{figure}

The optical properties of YNi$_2$B$_2$C have been studied several times
since the first measurement by Widder {\it et al.}\cite{widder},
who obtained anisotropic results
by linking together optical reflectance and electron energy
loss spectroscopy (EELS) measurements.
On the other hand, Bommeli {\it et al.} \cite{bommeli} reported that the
structural anisotropy would not affect the optical data based on the
observation that both the polycrystalline and single crystal samples
produced equivalent experimental results.
Therefore, the results were conflicting, and
until recently, the anisotropy problem in the optical properties of
Ni borocarbides has not been resolved\cite{jhkim,sjlee,mun}.
It is important to understand correct optical properties
since the superconducting energy gap is obtained from optical
measurements\cite{bommeli}.

Interestingly, all the above optical groups utilize the 3D electronic
structures of Ni borocarbides for their experiments.
It thus indicates that the electronic structure information only
may not be enough to determine definitely the optical anisotropy.
In this Letter, to resolve this controversial issue,
the optical conductivity and the electron energy loss spectroscopy (EELS)
have been studied theoretically for the first time
by using first principles calculations.
We have found that the optical properties of YNi$_2$B$_2$C are really 
anisotropic.

Calculations are performed
by employing the highly precise first principles
local density full-potential linearized augmented plane wave (FLAPW)
method\cite{flapw},
which makes no shape approximations to the charge density
or the potential.
The Hedin-Lundqvist form\cite{hedin} is employed for the LDA
exchange-correlation energy.
We used lattice parameters determined by experiment\cite{hong},
240 special ${\bf k}$-points\cite{kpts} for the self-consistency cycle
and 8619 ${\bf k}$-points for the calculation of the optical properties
inside the irreducible Brillouin zone.
The optical conductivity arises from the optical transitions of
the occupied valence electrons to unoccupied states above E$_F$\cite{mykim}.
Each interband transition is broadened in the calculation by introducing a
finite phenomenological inverse life time, $\Delta$=0.05 eV.
The Drude term due to free carriers is taken into account by
using parameters from experiment\cite{widder}.

Figure \ref{fig.bz} shows the crystal structure and
the first Brillouin zone of YNi$_2$B$_2$C.
In Fig. \ref{fig.bz}(b),
points a,b,c, and d are not symmetry points since the
symmetry is not a local maximum  at those points.
Points a and d (and b and c likewise) are the same points by the translational
symmetry in ${\bf k}$-space.
X has the highest symmetry as $\Gamma$ with 16 symmetry elements.

\begin{figure}[t]
     \epsfysize=10cm
     \epsfbox{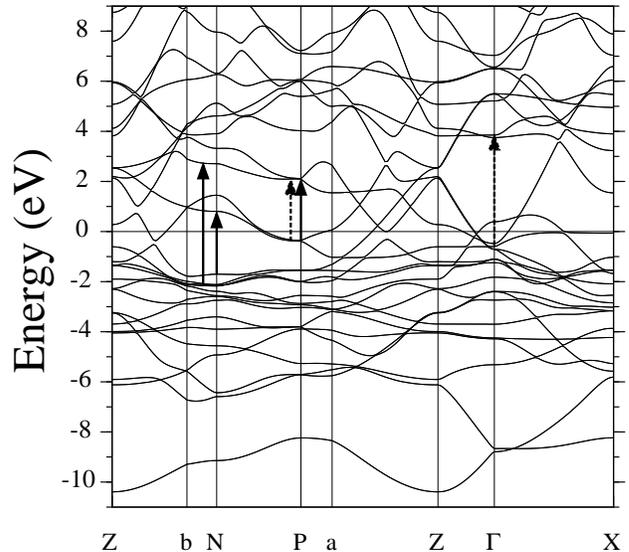}
     \caption{
     Electronic energy band structure of YNi$_2$B$_2$C.
     Dot and solid arrows indicate typical interband transitions
     along the $x$ and $z$ directions, respectively (see text and
	Fig. \ref{fig.gray}).
       \label{fig.band}
     }
\end{figure}

Figure \ref{fig.band} shows the calculated energy band structures
along the symmetry lines.
Band dispersions agree well with existing calculations.
However, in the present full-potential band calculation,
the 19th band at $\Gamma$, which gives a
small calabash shaped Fermi surface in the LMTO-ASA
calculation\cite{jilee},
is lifted up to higher energy.

\begin{figure}[t]
     \epsfysize=10cm
     \epsfbox{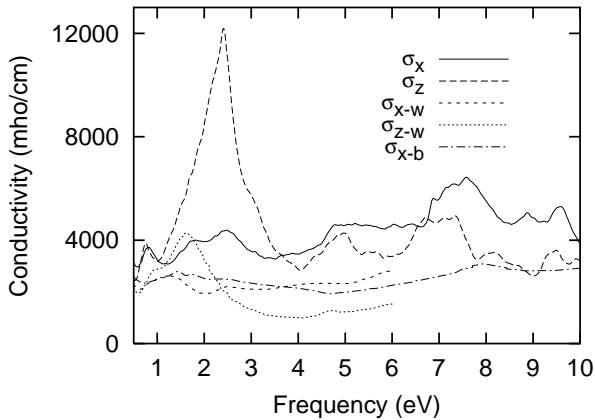}
     \caption{
     Optical conductivity of YNi$_2$B$_2$C.
     $\sigma_x$ and $\sigma_z$ represent results from the calculation.
     Experimental data are taken from
      Ref. [17]($\sigma_x$-w and $\sigma_z$-w) and Ref. [18]($\sigma_x$-b).
       \label{fig.sigma}
     }
\end{figure}

Figure \ref{fig.sigma} compares the calculated optical conductivities
for YNi$_2$B$_2$C with experiments,
in which $\sigma_x$ and $\sigma_z$ represent theoretical conductivities
along the $x$ and $z$ directions, respectively.
One can see a clear difference between
$\sigma_x$ and $\sigma_z$, implying the anisotropic nature of the
optical conductivity.
Anisotropic optical conductivities in the calculation are
consistent with the anisotropy observed in the experiment
by Widder {\it et al.} \cite{widder}, represented by
$\sigma_{x}$\_w and $\sigma_{z}$\_w in Fig. \ref{fig.sigma}.
Peaks are located at 2.48, 7.6, 8.9, and 9.6 eV for
$\sigma_x$, and at 2.4, 5.0, 8.16, and 9.5 eV for $\sigma_z$ in the 
calculation.
The most prominent and anisotropic peak in the calculation is the one at
2.4 eV of $\sigma_z$, while
$\sigma_x$ has a peak at 2.48 eV with a smaller magnitude than $\sigma_z$.
However, the intensity and position of the peak of $\sigma_z$ are somewhat
different from those in experiment: the highest peak is located at 1.6 eV,
which is lower by 0.8 eV than the theoretical one.
It is unusual that the experimental peak is located at lower frequency
considering that, in most LDA calculations, theoretical peaks are found
at lower frequencies than experimental ones.
The disagreement is to be clarified further theoretically or experimentally.
Other optical properties such as the reflectivity $R$ and
the dielectric constants $\epsilon$
show similar anisotropies which will be discussed elsewhere.

\begin{figure}[t]
     \epsfysize=10cm
     \epsfbox{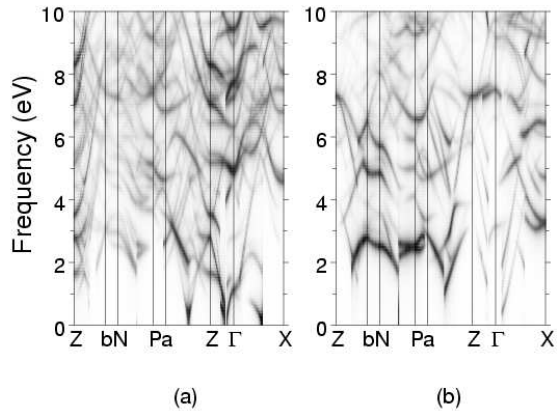}
     \caption{
     Momentum resolved optical conductivity $\sigma({\bf k},\omega)$
     for (a) $x$ and (b) $z$ direction, respectively.
       \label{fig.gray}
     }
\end{figure}

Bommeli {\it et al.} \cite{bommeli} obtained isotropic optical
conductivity (denoted by $\sigma_{x}$\_b in Fig. \ref{fig.sigma}).
The overall shape of $\sigma_{x}$\_b is seemingly close to
$\sigma_x$\_w, except that some fluctuations occur when
the relative magnitude of $\sigma_z$\_w with respect to $\sigma_{x}$\_w
becomes large.
This feature can be understood if $\sigma_{x}$\_b
corresponds to the average of the anisotropic optical conductivity.
Although it is not clear in $\sigma_{x}$\_b,
there is a peak at around the 2.4 eV in $\sigma_{x}$\_w.
The 2.4 eV peak in $\sigma_{x}$ is also observed in a recent experiment by
Lee {\it et al.}\cite{sjlee}.
Hence the experimental peak at 2.4 eV can be assigned to the calculated
$\sigma_x$ peak at 2.48 eV,

The optical conductivity is obtained directly from the optical
transitions between energy levels in the solid.
Hence the peaks in the optical conductivity can be assigned to
specific interband transitions.
However, peaks in derived quantities from the optical conductivity,
such as the reflectivity $R$ and the refractive index $n$,
may be shifted from the peaks of the optical conductivity.
In order to identify the peaks manifested in the optical conductivity,
we plot in Fig. \ref{fig.gray} the momentum resolved
optical conductivity $\sigma({\bf k}, \omega)$
along the symmetry lines in the Brillouin zone.
Integrating $\sigma({\bf k}, \omega)$  over
${\bf k}$-space gives rise to the optical conductivity $\sigma(\omega)$.

Figures \ref{fig.gray} (a) and (b) provide the optical transitions along
the $x$ and $z$ directions, respectively.
The darker intensity in Fig. \ref{fig.gray} represents the stronger
interband transitions.
It is evident from Figs. \ref{fig.gray}(a) and (b) that
the optical transitions are anisotropic between the $x$ and
$z$ directions.
The anisotropy is due to different matrix element between the $x$ and $z$
directions.
Strong transitions in the $z$-direction occur around P and N,
while they occur at points near $\Gamma$ in the $x$-direction.
The anisotropy is most prominent at the 2.4 eV transition near P;
there is a strong transition in the $z$-direction,
but a negligible transition in the $x$-direction.
The intensity of the 2.4 eV transition is further enhanced by
the transition near N.
In Fig. \ref{fig.band}, corresponding transitions for 2.4 eV
are marked as small solid arrows at N and P.
Other strong transitions identified in Fig. \ref{fig.gray}
are also marked as arrows in Fig. \ref{fig.band}.
It is noteworthy that
the unoccupied bands (19th band) for the 2.4 eV transition near P
have similar shapes to the occupied bands (18th band), reminiscent of
the nesting of bands.

In order to see the detailed nature of the transition,
we have calculated contributions from each atomic species
for the bands related to the 2.4 eV transition at P.
The 18th band consists of Ni-$d_{3z^2-r^2}$(36\%) and Ni-$d_{xy}$(35\%)
states, while the unoccupied band (19th band) consists of
Ni-$d_{xy}$(16\%), Ni-$s$(12\%), Ni-$p_z$(4.4\%), B-$p$(6.9\%), and
C-$p$(16\%) states.
Since the occupied band originates mainly from Ni-$d$ states,
the transition matrix element is determined by Ni site.
By the selection rule about the symmetry of the wave functions,
one can assign the 2.4 eV peak in $\sigma_z$ to
the transition between Ni-$d_{3z^2-r^2}$ and Ni-$p_z$.
The transition matrix element for $\sigma_x$ is zero in this case,
which explains
the negligible intensity at 2.4 eV near P in $\sigma_x({\bf k}, \omega)$.
Thus Ni states play a vital role in yielding
a large anisotropic peak in $\sigma_z$.
This is indeed consistent with the dominant Ni-$d$ DOS near the Fermi level
\cite{jilee}.

\begin{figure}[t]
     \epsfysize=10cm
     \epsfbox{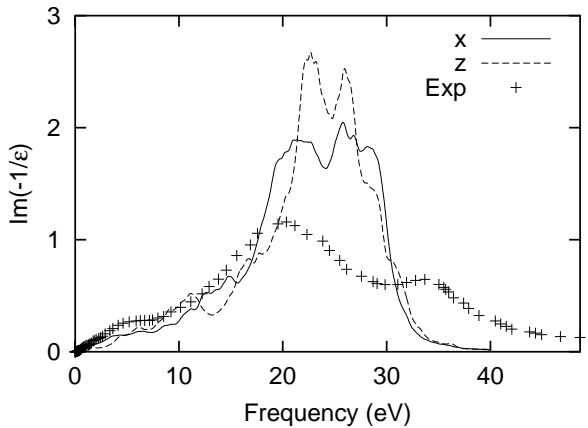}
     \caption{
     Electron energy loss (EELS) spectrum Im($-1/\epsilon$) of YNi$_2$B$_2$C.
     Crosses represent experimental data reproduced from the experimental
     Widder {\it et al.} [17].
       \label{fig.eels}
     }
\end{figure}

Figure \ref{fig.eels}\ depicts the loss function,
Im$(-1/\epsilon(\omega))$, of the EELS measurement with zero momentum transfer.
There is no EELS data available for single crystals.
Hence, in the figure, the calculated result is compared with
Widder {\it et al.}'s experiment for polycrystals \cite{widder}.
There is no plasma peak due to free carriers which is expected
to occur between 3.75 eV and 4.25 eV\cite{widder,jhkim,sjlee}.
It is damped due to interband transitions (Landau damping),
in agreement with the experimental EELS result.
The calculated loss function shows a two peak structure in both the
$x$ and $z$-direction,
and the spectrum in the $z$-direction is narrower and higher
than that of $x$ direction.
Two peaks are located at 22.7 eV and 26 eV in the z-direction.
In YNi$_2$B$_2$C, there are 33 valence electrons per formula unit.
The simple plasma frequency relation, $\omega_p^2=4\pi ne^2/m$,
gives 26.3 eV which agrees well with the above plasma frequency
of 26 eV.
Therefore, one can explain the plasma frequency at 26 eV as originating
from the response of all valence electrons to the external
electric field.
Then the peak at 22.7 eV can be ascribed to the valence electrons near E$_F$:
Ni-$d$, Y-$d$, B-$p$, and C-$p$ states.
In the experiment, two peaks are observed at $\sim$ 20 eV and $\sim$ 35 eV.
Although the two peak structure in the calculation
agrees qualitatively with experiment, the peak positions are different.
The EELS experiment with single crystal is required to
clarify the discrepancy between theory and experiment.

In conclusion, we have investigated the optical properties of YNi$_2$B$_2$C
by using the first-principles FLAPW method.
The optical conductivity is found to be anisotropic between
the $x$ and the $z$ direction.
A strong peak at 2.4 eV is obtained in the $z$ direction, which
is assigned to the transition near P
between Ni-$d_{3z^2-r^2}$ and Ni-$p_z$ states,
indicating that the Ni-site plays an important role in the anisotropy
of YNi$_2$B$_2$C.
The calculated EELS spectra are also found to be anisotropic
with a two peak structure.

\acknowledgements
The authors are grateful to KISTEP/CRI of Korean ministry of
Science and Technology for financial support.
BIM is also supported by the KOSEF through the eSSC at POSTECH.
SJY thanks to Prof. J. H. Kim for helpful discussions. Work at
Northwestern University supported by the U.S. Department of Energy
(Grant No. DE-F602-88ER45372).



\begin{thebibliography}{100}
%
\bibitem{nagarajan} R. Nagarajan {\it et al.},
     Phys. Rev. Lett. {\bf 72}, 274 (1994).
\bibitem{cava} R. J. Cava {\it et al.}, Nature {\bf 367}, 252 (1994).
\bibitem{canfield} P. C. Canfield, P. L. Gammel, and D. J. Bishop,
     Phys. Today {\bf 51} (10), 40 (1998).
\bibitem{mattheiss} L. F. Mattheiss, Phys. Rev. B {\bf 49}, 13 279 (1994).
\bibitem{pickett} W. E. Pickett and D. J. Singh, Phys. Rev. Lett. {\bf
72}, 3702 (1994).
\bibitem{jilee} J. I. Lee {\it et al.}, Phys. Rev. B {\bf 50}, 4030 (1994).
\bibitem{singh} D. J. Singh, Solid State Comm. {\bf 98}, 899 (1996).
\bibitem{civale} L. Civale {\it et al.}, Phys. Rev. Lett. {\bf 83},
3920 (1999).
\bibitem{johnston} E. Johnston-Halperin {\it et al.}, Phys. Rev. B
{\bf 51}, 12852 (1995).
\bibitem{fisher} I. R. Fisher, J. R. Cooper, and P. C. Canfield,
       Phys. Rev. B {\bf 56}, 10 820 (1997).
\bibitem{rathnayaka} K. D. D. Rathnayaka {\it et al.},
      Phys. Rev. B {\bf 55}, 8506 (1997).
\bibitem{vaglio} R. Vaglio {\it et al.}, Phys. Rev. B {\bf 56}, 934 (1997).
\bibitem{sera} M. Sera {\it et al.}, Phys. Rev. B {\bf 54}, 3062 (1996).
\bibitem{sakata} H. Sakata {\it et al.}, Phys. Rev. Lett. {\bf 84}, 
1583 (2000).
\bibitem{kumagai} K. Kumagai {\it et al.}, J. Low Temp. Phys. {\bf
105}, 1641 (1996).
\bibitem{vonlips} H. von Lips {\it et al.}, Phys. Rev. B {\bf 60}, 11
444 (1999).
\bibitem{widder} K. Widder {\it et al.}, Europhys. Lett {\bf 30}, 55
(1995); K. Widder {\it et al.}, J. Low Temp. Phys. {\bf 105}, 516
(1996).
\bibitem{bommeli} F. Bommeli {\it et al.}, Phys. Rev. Lett. {\bf 78},
547 (1997).
\bibitem{jhkim} J. H. Kim {\it et al.}, Physica C {\bf 341}, 2233 (2000).
\bibitem{sjlee} S. J. Lee, B. K. Cho, P. C. Canfield, and D. W. Lynch,
    Phys. Rev. B {\bf 63}, 233103 (2001).
\bibitem{mun} M.-O. Mun {\it et al.}, J. Kor. Phys. Soc. {\bf 39}, 406 (2001).
\bibitem{flapw} E. Wimmer, H. Krakauer, M. Weinert, and
     A. J. Freeman, Phys. Rev. B {\bf 24}, 864(1981);
     M. Weinert, E. Wimmer, and A. J. Freeman, {\it ibid.} {\bf 26}, 4571(1982).
\bibitem{hedin} L. Hedin and B. I. Lundqvist, J. Phys. C. {\bf 4}, 2064 (1971).
\bibitem{hong} N. M. Hong {\it et al.},
      Physica C {\bf 227}, 85 (1994).
\bibitem{kpts} H. J. Monkhorst and J. D. Pack, Phys. Rev. B {\bf 13},
5188 (1976).
\bibitem{mykim} M. Kim, A. J. Freeman, and R. Wu, Phys. Rev. B {\bf
59}, 9432 (1999).

\end{thebibliography}
\end{document}